\documentstyle[12pt]{article}
\footskip- 15mm
\begin{document}
\begin{center}
Marginal hyperchaos synchronization with a single driving variable\\
ELMAN M. SHAHVERDIEV,\footnote{e-mail:shahverdiev@lan.ab.az}\\
Institute of Physics of Academy of Sciences of Azerbaijan,\\
33,H.Javid Av., 370143-Baku,Azerbaijan.\\
\end{center}
The seminal papers by Pecora and Carrol (PC) [1]  and Ott, Grebogi 
and Yorke (OGY) [2] in 1990 have induced avalanche of research works 
in the field of chaos control. Chaos synchronization in dynamical 
systems is one of methods of controling chaos, see, e.g. [1-8] and 
references therein.The interest to chaos synchronization in part is 
due to the application of this phenomenen in secure communications, 
in modeling of brain activity and recognition processes,etc [1-8]. 
Also it should be mentioned that this method of chaos control may 
result in improved performance of chaotic systems [1-8].
According to PC [1] synchronization of two systems occurs when the 
trajectories of one of the systems will converge to the same values as 
the other and they will remain in step with each other. For the 
chaotic systems synchronization is performed by the linking of 
chaotic systems with a common signal or signals (the so-called 
drivers): suppose that we have a chaotic dynamical system of 
three or more state variables. In the above mentioned way of chaos 
control one or some of these state variables can be used as an input 
to drive a subsystem consisting of remaining state variables and which 
is a replica of part of the original system.In [1] it has been shown 
that if the real parts of the Lyapunov exponents for the subsystem 
(below: sub-Lyapunov exponents) are negative then the subsystem 
synchronizes to the chaotic evolution of original system. If the 
largest sub-Lyapunov exponent is not negative, then one can use the 
nonreplica approach to chaos synchronization [9]. Within the 
nonreplica approach to chaos synchronization one can try to perform 
chaos synchronization between the original chaotic system and 
nonreplica response system with control terms vanishable upon 
synchronization.To be more specific, one can try to make negative the 
real parts of the conditional Lyapunov exponents of the nonreplica 
response system. As it has been shown in [9] from the application 
viewpoint nonreplica approach has some advantages over the replica 
one. \\ 
Recently in [10] it has been indicated that for more secure
communication purposes the use of hyperchaos is more reliable. Quite 
naturally in the light of this result the investigation of hyperchaos
synchronization is of paramount importance. According to Pyragas for 
hyperchaos synchronization at least two drive variables are needed 
[11].\\
Recently this idea was challenged in [12] in the sense that instead 
of several driving variables one can try to drive the response system
with a scalar combination of those driving variables. But one should 
keep in mind that in this case the synchronization occurs between the 
nonreplica system and original chaotic system.Recent paper [13] also 
falls into this category, although its authors are using only single 
control term added to the replica response system.\\
In recent work [14] the classification of different types of 
synchronization is conducted. Such a classifiation into different 
types corresponds to the different values for the sub- (or 
conditional) Lyapunov exponents and still there is no unique generally 
accepted classification. For example, according to [15] if one of sub-
Lyapunov exponents is equal to zero, while others are negative, then 
one can still speak of synchronization between the response and drive 
systems in the general sense: a generalized synchronization
introduced for drive-response systems is defined as the presence of 
some functional relation between the states of response and drive.  
According to [14], the similar situation could be characterized by 
the so-called marginal synchronization:there are there types of 
marginal synchronization:1) marginal constant synchronization: in 
this case the response system becomes synchronized with the drive, 
but with a constant separation.\\
2) marginal oscillatory synchronization: this type of synchronization 
implies that the difference between the drive and response will change 
in an oscillatory fashion with a frequency that will depend on the 
imaginary part and with constant amplitude that will be related to 
the difference at the moment in  which the connection starts.\\
3) sized synchronization: in this type of synchronization also one 
has a single zero sub-Lyapunov exponent; in this case the observed 
behavior is different from the case of marginal constant 
synchronization and consists in that the response system exhibits the 
same qualitative behavior as the drive, but with different size (and 
sometimes with different symmetry); as a prominent example of this 
type of synchronization one can cite the case $z$ driving for the 
classical Lorenz model [7]. It is easy to show that  one of sub-
Lyapunov exponents for the Lorenz model in the case of $z$ driving 
really is equal to zero. By simple calculations one can easily obtain 
the following equation for the sub-Lyapunov exponents:
$$\lambda ^{2} + \lambda (\sigma +1) $$
$$- \sigma (r-z)=0,\hspace*{3cm}(1)$$
where $z=z(t)$ is the solution of the original Lorenz system. As it 
has been shown in [16], for those dynamical systems, whose chaotic 
behavior has arisen out of instability of the steady state solutions 
(fixed points) while calculating the sub-Lyapunov exponents one can 
replace the time dependent solutions of the dynamical systems with the 
steady state (st) solutions safely. As the Lorenz model has the above-
mentioned property, and $z^{st}=r-1$, one can easily establish that 
one of the sub-Lyapunov exponents is equal to zero.\\
In the above mentioned papers [14-15] the presented examples represent
third-order nonlinear dynamical systems.\\
In this paper we present an example of marginal or general type 
synchronization in higher dimensional system, to be more specific in 
one of four dimensional hyperchaos R\"ossler models:
$$\frac{dx}{dt}= - y - z - w, $$
$$\frac{dy}{dt}= x, \hspace*{3cm}(2)$$
$$\frac{dz}{dt}= a (y - y^{2}) - bz, $$
$$\frac{dw}{dt}= c (\frac{z}{2} - z^{2}) - dw, $$
According to [17,18], nonlinear system (1) exhibits hyperchaotic 
behavior with some positive values of system's parameters $a, b, c, 
d$. First consider as a driver state variable $x$. Then the response 
system could be written in the following form:
$$\frac{dy_{r}}{dt}= x =A_{1},$$
$$\frac{dz_{r}}{dt}= a (y_{r} - y_{r}^{2})$$  
$$-bz_{r}=A_{2},\hspace*{3cm}(3) $$
$$\frac{dw_{r}}{dt}= c (\frac{z_{r}}{2} - z_{r}^{2})$$
$$ - dw_{r}=A_{3},$$
The eigenvalues of the Jacobian of (3) are to be found from the 
equation: 
$$\lambda (\lambda + b)(\lambda + d)=0, \hspace*{3cm}(4)$$
In other words, in the case of $x$ driving, according to 
classification of [14] marginal constant synchronization is possible, 
as one of sub-Lyapunov exponents is negative, while others -positive.
 Surprisingly, due to the form of the nonlinear system under study in 
the case of $y$ driving we obtain exactly the same equation for the 
sub-Lyapunov exponents. So the marginal constant synchronization takes 
place in the case of $y$ driving too.\\
As the investigations show quite different type of marginal 
synchronization, namely marginal oscillatory synchronization could be 
realized in the case of $z$ driving. Indeed, as the calculations 
indicate in this case the conditional Lyapunov exponents satisfy the 
equation: 
$$(\lambda +d) (\lambda ^{2} + 1) = 0,\hspace*{6cm}(5)$$
According to the classification in [14], this case quite "eligible" 
to be named as the marginal oscillatory synchronization, as one of sub-
Lyapunov exponents is negative, while the two others are complex 
conjugate with zero real parts.So far we considered the cases of 
driving with $x, y, z$ variables and we have succeeded in marginal 
synchronization of hyperchaos only using the fact of positiveness of 
part of the system's parameters, namely $b, d$. As the calculations 
show the case of $w$ driving is a bit more complicated in the 
sense that some additional relationships between the system's 
parameters are required. Depending on these relationships different 
types of synchronization, according to the classification of [14] are 
also possible.\\
Thus in this paper for the first time (to our knowledge) we have 
demonstrated the possibility of hyperchaos synchronization with a 
single driving variable within the replica approach. 

\end{document}